\documentclass[prl,twocolumn,showpacs,amsmath,amssymb,superscriptaddress]{revtex4-1}
\usepackage{graphicx}
\usepackage{dcolumn}
\usepackage{bm}
\usepackage{bbm}
\usepackage{ulem}
\usepackage{color, soul}
\usepackage{epstopdf}
\usepackage{amssymb}
\usepackage{amsmath}

\begin{document}

\title{Observation of Photonic Topological Valley-Hall Edge States}

\author{Jiho Noh$^{1*}$, Sheng Huang$^{2*}$, Kevin Chen$^{2}$, and Mikael C. Rechtsman$^{1}$\\
  {{\it 
$^{1}$Department of Physics, The Pennsylvania State University, University Park, PA 16802, USA\\$^{2}$Department of Electrical and Computer Engineering, University of Pittsburgh, Pittsburgh, PA 15261, USA\\
}}}

\begin{abstract}
We experimentally demonstrate topological edge states arising from the valley-Hall effect in two-dimensional honeycomb photonic lattices with broken inversion symmetry. We break inversion symmetry by detuning the refractive indices of the two honeycomb sublattices, giving rise to a boron nitride-like band structure. The edge states therefore exist along the domain walls between regions of opposite valley Chern numbers. We probe both the armchair and zig-zag domain walls and show that the former become gapped for any detuning, whereas the latter remain ungapped until a cutoff is reached. The valley-Hall effect provides a new mechanism for the realization of time-reversal invariant photonic topological insulators.
\end{abstract}

\maketitle
Photonic topological insulators (PTIs) are dielectric structures that possess topologically protected edge states that are robust to scattering by disorder \cite{Lu2014,Haldane2008,Wang2009,Carusotto2011,Hafezi2011,Fang2012,Rechtsman2013,Hafezi2013,Khanikaev2013,cheng2016robust,Wu2015,Gao2016}. There are two categories of PTIs: those that break time-reversal symmetry \cite{Wang2009,Rechtsman2013} and those that preserve it \cite{Hafezi2013,Khanikaev2013,Wu2015}.
In PTIs that break time-reversal symmetry, there exist one-way edge states, which ensure their robustness, due to the lack of counter-propagating partners at same frequency.
In those that preserve it, there exist counter-propagating edge states that are protected only against certain classes of disorder.  However, the latter can be more straightforward to realize because they do not require strong time-reversal breaking. Photonic topological insulators have been of interest due to the possibility of photonic devices that are less sensitive to fabrication disorder.

In the valley-Hall effect, broken inversion symmetry in a two-dimensional honeycomb lattice causes opposite Berry curvatures in the two valleys of the band structure \cite{Xiao2007,Yao2008}, and has been realized in solid-state two-dimensional materials \cite{Mak2014,Gorbachev2014,Sui2015,Shimazaki2015,Ju2015}.
The valley-Hall effect is time-reversal invariant and has common characteristics with the spin Hall effect \cite{Kane95}, where the two valleys in the band structure are used as `pseudo-spin' degrees of freedom. 
It was shown theoretically that valley-Hall topological edge states would arise in analogous photonic structures \cite{Ma2016,Dong2017,Chen2016_01,Chen2016_02,fefferman2016bifurcations,fefferman2016edge,fefferman2017}.
In addition, valley-Hall topological edge states have also been recently studied in the context of topological valley transport of sound in sonic crystals \cite{Lu2017}.

Here, we present the experimental observation of photonic topological valley-Hall edge states at domain walls between valley-Hall PTIs of opposite valley Chern numbers.
The bulk-edge correspondence ensures the presence of edge states: the change in valley Chern number across the domain wall is associated with the existence of counter-propagating edge states \cite{Mong2011,Delplace2011,Ju2015}.
We realize the photonic valley-Hall topological edge states in evanescently-coupled waveguide arrays, i.e., photonic lattices, fabricated using the femtosecond direct laser writing technique \cite{Szameit2010}.
We probe different types of domain walls, namely the armchair and zig-zag edges. We also enter a fully gapped regime, which is not accessible in solid-state two-dimensional materials.
The topological protection associated with the valley-Hall effect applies as long as a single valley is populated and does not mix with the other valley. In general, disorder that has only low spatial frequency components (i.e., is sufficiently smooth) will not mix the valleys.
\begin{figure*}[htbp]
	\centering
	\includegraphics[width=17.0cm]{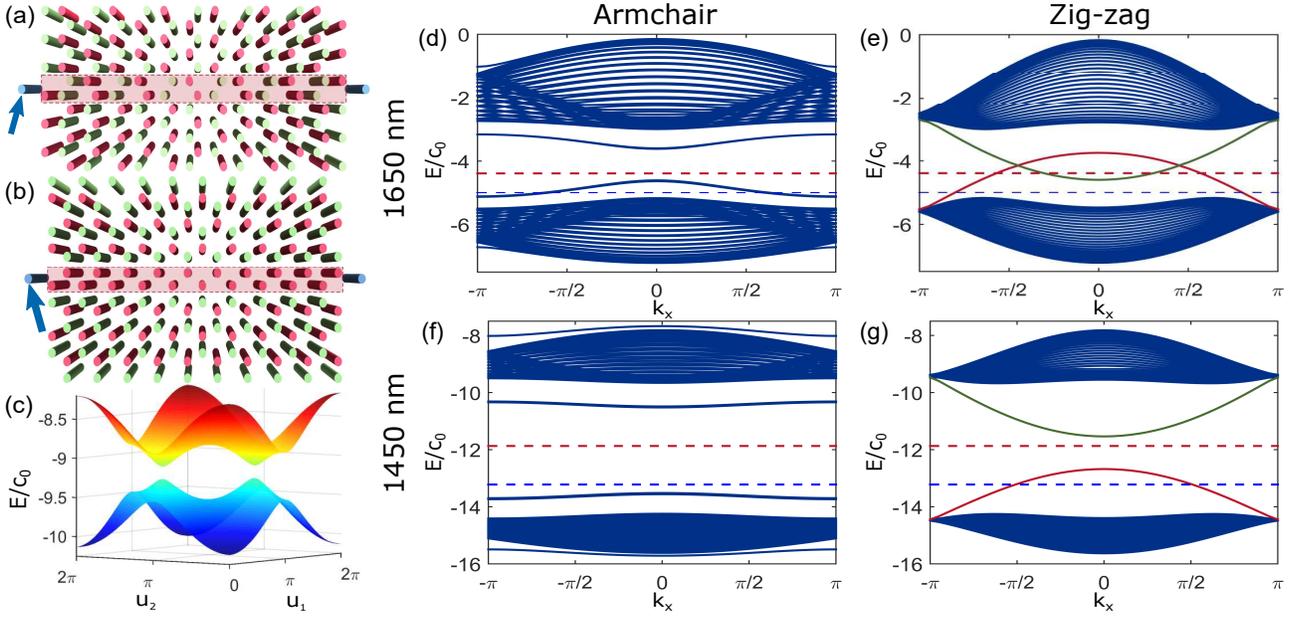}
	\caption{(a) Schematic diagram of inversion-symmetry-broken honeycomb lattices with armchair and (b) and zig-zag edge domain walls. Red and green waveguides indicate different refractive index, and blue indicates straw waveguides. Red shaded regions indicate domain walls. (c) Band structure of the inversion-symmetry-broken graphene defined by $u_{1}\textbf{\textrm{b}}_{1}+u_{2}\textbf{\textrm{b}}_{2}$, where $\textbf{\textrm{b}}_{1}=\frac{2\pi}{3a}\left(1,\sqrt{3}\right)$, $\textbf{\textrm{b}}_{2}=\frac{2\pi}{3a}\left(1,-\sqrt{3}\right)$ are the reciprocal lattice vectors and $a$ is the lattice constant. (d) Continuum edge band structures with periodic boundary conditions on both $x$ and $y$ directions at $\lambda$=1650nm and $k_{y}$=0 when the armchair or (e) zig-zag edges are placed at the domain wall. (f-g) Corresponding band structures at $\lambda$=1450nm. Red and blue dashed lines indicate the energy of eigenmodes that are excited by coupling with the straw waveguide when we excited modes at mid-gap and significantly below mid-gap, respectively.  Green and red bands in (e,g) indicate edge states located at the domain wall close to and far away from the straw waveguide, respectively. Therefore, only the green bands are accessible in the experiment by exciting either straw waveguide.}
	\label{Fig_01}
\end{figure*}

We begin by describing our experimental system, which is composed of an array of evanescently-coupled waveguides arranged in a honeycomb lattice geometry.
The laser-writing technique allows us to arbitrarily control the refractive index of the waveguides, by varying the average power of the pulse train in the femtosecond direct laser writing procedure.
The geometries of lattices having armchair and zig-zag edges at their domain wall are depicted in Fig. \ref{Fig_01}(a) and (b), respectively.
The interface is between two regions (top and bottom) that are both honeycomb lattices with opposite signs of the on-site energy detuning between the two component sublattices (this breaks inversion symmetry within each given lattice).
The detuning is shown in the figure by the different colors (red and green) of the component sublattices.
Experimentally, the detuning is carried out by controlling the refractive index of the waveguide at each site.
Fig. \ref{Fig_01}(c) shows the two-dimensional bulk band structure of the inversion-symmetry-broken honeycomb lattice, clearly showing the two valleys.
This is simply an inversion-symmetry-broken variation of the photonic honeycomb lattices \cite{Peleg2007,Bahat-Treidel2008,Plotnik2013}, and as in the graphene band structure, two valleys are located at two non-equivalent K and K$^{\prime}$ points in the first Brillouin zone.
The valley Chern number is defined as the difference in integrated Berry curvature associated with the two valleys.
Since the Berry curvature points in the opposite directions ($+z,-z$) in the two valleys in a given lattice, and the sign is given by that of the inversion-breaking term, it follows that the top and bottom lattices in Fig. \ref{Fig_01}(a,b) must have opposite valley Chern numbers and thus have valley-protected edge states.  

The diffraction of light through the waveguide array is governed by the paraxial wave equation:
\begin{equation}
\begin{split}
i\partial_{z}\psi(\textbf{\textrm{r}},z) = -\frac{1}{2k_{0}}\nabla^{2}_{\textbf{\textrm{r}}}\psi(\textbf{\textrm{r}},z)-\frac{k_{0}\Delta n(\textbf{\textrm{r}})}{n_{0}}\psi (\textbf{\textrm{r}},z)\\
\equiv H_{cont}\psi(\textbf{\textrm{r}},z),
\label{eq_paraxial}
\end{split}
\end{equation}
where $\psi(\textbf{\textrm{r}},z)$ is the envelope function of the electric field $\textrm{\textbf{E}}(\textbf{\textrm{r}},z)=\psi(\textbf{\textrm{r}},z)\exp^{i(k_{0}z-\omega t)}\hat{x}$, $k_{0}=2\pi n_{0}/\lambda$ is the wavenumber within the medium, $\lambda$ is the wavelength of the laser light, $\omega = 2\pi c/\lambda$, and $\nabla^{2}_{\textbf{\textrm{r}}}$ is the Laplacian in the transverse $(x,y)$ plane. $H_{cont}$ is the continuum Hamiltonian for propagation of the wave in the photonic lattice.
$\Delta n$ is the refractive index of the waveguide relative to the index of our medium, $n_{0}=1.47$, which acts as an effective potential in the Schr\"{o}dinger equation, Eq. (\ref{eq_paraxial}).
The inversion symmetry of the lattice is broken by having different $\Delta n_{\textrm{A}}$ and $\Delta n_{\textrm{B}}$ for waveguides in sublattices $A$ and $B$, respectively, which is analogous to having different on-site energies $E_{\textrm{A}}$ and $E_{\textrm{B}}$ in the condensed-matter context.
Furthermore, we write two additional waveguides, which we call `straw waveguides' (as discussed previously in Ref. \cite{CleoFTAI}) into which light is injected.
The straws are weakly coupled to the lattice, allowing them to act as an external drive that is injecting light into the system without altering the system's intrinsic modes.
Furthermore, varying the refractive index of the straw, $\Delta n_{\textrm{s}}$, allows for the control of the propagation constant (i.e., energy) of the modes being injected into the structure.
By analogy with condensed-matter systems, the straw allows us to control the effective `Fermi energy' of the system, only allowing coupling to modes of a given energy, $E$.

The emergence of valley-Hall topological edge states is shown by a full-continuum calculation by diagonalizing $H_{cont}$ in Eq. (\ref{eq_paraxial}) of two-dimensional inversion-symmetry-broken honeycomb lattice ribbons.
The unit cell is a strip that is periodic in the horizontal direction (with a periodicity given by the lattice constant), but is many unit cells in the vertical direction and includes the domain wall (in fact, it must contain a minimum of two domain walls).
The eigenvalues of the Schr\"odinger operator given in Eq. (1) are the energies of the calculated eigenmodes.
Band structures and therefore bandgap sizes can be engineered by sweeping across $\Delta E/c_{0}$, where $\Delta E = E_A-E_B$ is the difference between the on-site energies in the two sublattices and $c_{0}$ is the coupling strength between the nearest-neighbor waveguides.
Experimentally, $\Delta E$ can be controlled by varying both $\Delta n_{\textrm{A}}$ and $\Delta n_{\textrm{B}}$, and $c_{0}$ can be increased by decreasing the distance between the nearest-neighbor waveguides, $d$, and increasing $\lambda$; $c_0(\lambda)$ at fixed $d$=$19\,\mu$m for $\lambda$=1650nm and $\lambda$=1450nm are 2.69cm$^{-1}$ and 1.76cm$^{-1}$, respectively (for the remainder of the work, we logically order long wavelength before short wavelength because the bandgap increases with decreasing wavelength).
Fig. \ref{eq_paraxial}(d) and (e) show band structures when the armchair and zig-zag edges are placed at the domain wall, respectively, where $\lambda$=1650nm, $d$=$19\,\mu$m, $\Delta n_{\textrm{A}}=2.50\times 10^{-3}$, and $\Delta n_{\textrm{B}}=2.90\times 10^{-3}$, and Fig. \ref{eq_paraxial}(f,g) show corresponding band structures with same $d$, $\Delta n_{\textrm{A}}$ and $\Delta n_{\textrm{B}}$ but with $\lambda$=1450nm.
For both structures with armchair and zig-zag edge domain walls, the bulk bandgap opens immediately as $\Delta E/c_{0}$ becomes nonzero.
However, behaviors of the edge states are different for each case: for the structure with the armchair domain wall, the edge bandgap opens immediately after $\Delta E/c_{0}$ becomes nonzero (Fig. \ref{eq_paraxial}(d,f)).
For the structure with the zig-zag domain wall, there exists edge states at the mid-gap for small $\Delta E/c_{0}$ (Fig. \ref{eq_paraxial}(e)), which indicates the edge bandgap would open only at finite $\Delta E/c_{0}$ (Fig. \ref{eq_paraxial}(g)).
The two edge state bands shown in green and red in Fig. \ref{eq_paraxial}(e,g) are localized close to the straw waveguides (in the center of the figure), and far away from them, respectively.
Therefore, only the green bands will be physically accessible in the experiment.  
This difference between the armchair and zig-zag edges arises because the orientation of the armchair termination is such that it mixes the two valleys; since they may scatter between them, this allows for a matrix element for a gap to open even for small $\Delta E/c_0$.
However, the zig-zag edge runs parallel to the line that connects the two valleys in $k$-space, implying that the presence of the edge does not connect them, allowing them to remain ungapped.  

\begin{figure}[htbp]
	\centering
	\includegraphics[width=8.0cm]{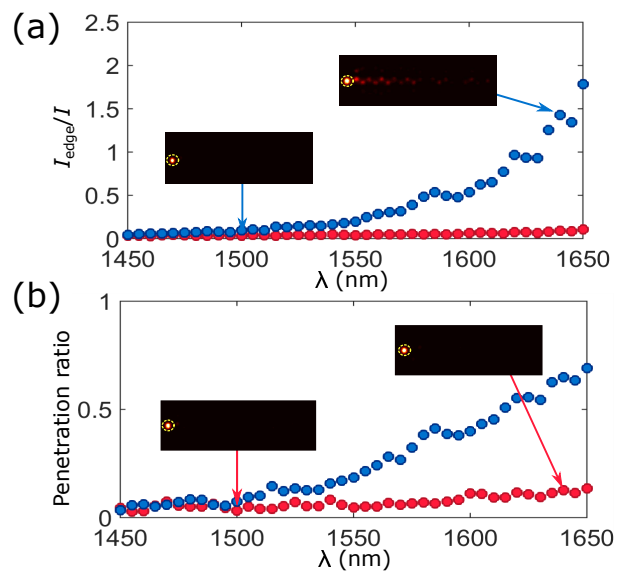}
	\caption{(a) Measured edge intensity ratio and (b) penetration ratio when we excite modes at mid-gap. Blue and red dots are from zig-zag and armchair edge domain walls, respectively. (inset) Diffracted light measured at the output facet. The waveguide where light is initially injected is marked with yellow dashed circle.}
	\label{Fig_02}
\end{figure}

To experimentally observe the emergence of topological edge states, a beam was launched at the input facet through a lens-tipped fiber, which allows us to couple the beam precisely into a selected straw waveguide.
Here, the refractive index of the straw waveguides was calibrated to inject light at the mid-gap and significantly below the mid-gap, in different devices.
The energies of the straw waveguide modes were calculated by diagonalizing $H_{cont}$ of a single waveguide.
In Fig. \ref{Fig_02}, we present the observed diffracted light at the output facet of the array for the case of mid-gap driving (red-dashed lines in Fig. \ref{Fig_01}(d-g)).
Here, the calculated energy of the straw waveguide modes at the mid-gap energy were -$4.39c_{0}$ and -$11.87c_{0}$ for 1650nm and 1450nm, respectively.
We plot: (1) the edge intensity ratio, i.e., the ratio of the light intensity along the domain wall ($I_{edge}$) to the total light intensity ($I$), and (2) the penetration ratio, which is the light intensity that penetrates into the structure normalized by that in the straw. 
First, in the photonic lattice with the armchair domain wall, we observe that most of the light coupled into the straw waveguide stayed in the straw, not coupling into the waveguide array (Fig. \ref{Fig_02} insets).
Both measured edge intensity ratio and penetration ratio were relatively very small, which indicate the presence of the bandgap between the edge modes: i.e., no edge states are available to transport light through the array. This experimental result agrees with the full-continuum calculation having red-dashed lines not crossing any edge states in the band structure as shown in Fig. \ref{Fig_01}(d,f).
On the other hand, from the analogous structure with zig-zag domain wall, we observed a clear excitation of edge states along the domain wall, which becomes more significant as wavelength is increased.
This indicates that at $\lambda$=1450nm, the bandgap is fully open so that the straw waveguide mode is not able to couple into the domain wall but as we increase $\lambda$ to make $\Delta E/c_{0}$ subsequently decrease, the bandgap becomes smaller and eventually edge states couple with the straw waveguide mode at mid-gap.
Furthermore, the sharp increase in edge intensity ratio and penetration ratio indicates that the edge state is topological and that there exist edge states having mid-gap energy, respectively.  This experimentally establishes the presence of valley-Hall edge states at mid-gap for the zig-zag edge, and the lack thereof for the armchair edge, consistent with theoretical predictions described above.

\begin{figure}[htbp]
	\centering
	\includegraphics[width=8.0cm]{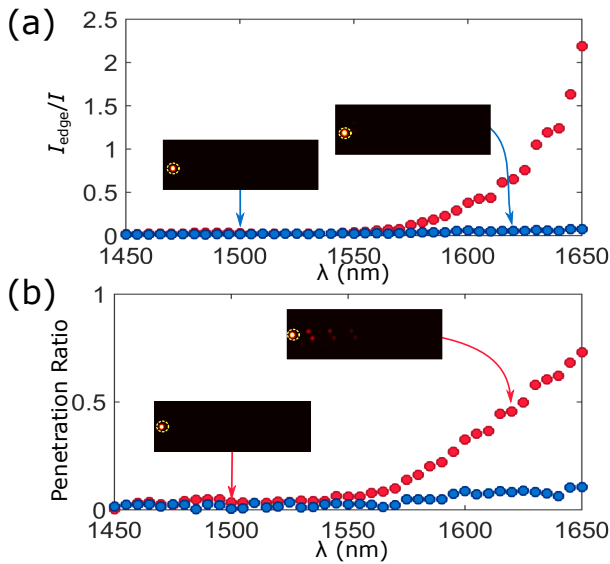}
	\caption{(a) Measured edge intensity ratio and (b) penetration ratio when we excited modes significantly below mid-gap. Blue and red dots are measured zig-zag and armchair edge domain walls, respectively. (inset) Diffracted light measured at the output facet. The waveguide where light is initially injected is marked with yellow dashed circle.}
	\label{Fig_03}
\end{figure}

We further probe the valley-Hall edge states by changing $\Delta n_{\textrm{s}}$ of the straw waveguides, while keeping $\Delta n_{\textrm{A}}$ and $\Delta n_{\textrm{B}}$ the same, such that we excite modes at a different energy (blue-dashed lines in Fig. \ref{Fig_01}(d-g)).
Here, the calculated energies of the straw waveguide modes were $-4.98 c_{0}$ and $-13.22 c_{0}$ for 1650nm and 1450nm, respectively.
For the armchair edge, the energy coincides with edge bands at $\lambda$=1650nm, but not at $\lambda$=1450nm (Fig. \ref{Fig_01}(d,f)).  
Therefore, we observe confinement to the input straw waveguide at $\lambda$=1450nm followed by increased penetration along the domain wall with increasing wavelength and strong penetration by $\lambda$=1650nm.  For the zig-zag edge, however, the energy does not coincide with the state along the domain wall boundary depicted in Fig. \ref{Fig_01}(b) (and whose dispersion is shown in green in Fig. \ref{Fig_01}(e,g)), but rather the confined state that arises on the opposite side of the system when periodic boundary conditions are imposed in the vertical direction (shown in red in Fig. \ref{Fig_01}(e,g)).  In other words, since the only edge states localized near the straw waveguide are those drawn in green, there is no penetration along the zig-zag edge for this energy.  Therefore, no penetration is observed in the entire wavelength range for the zig-zag edge (Fig. \ref{Fig_03}).   

\begin{figure}[htbp]
	\centering
	\includegraphics[width=8.0cm]{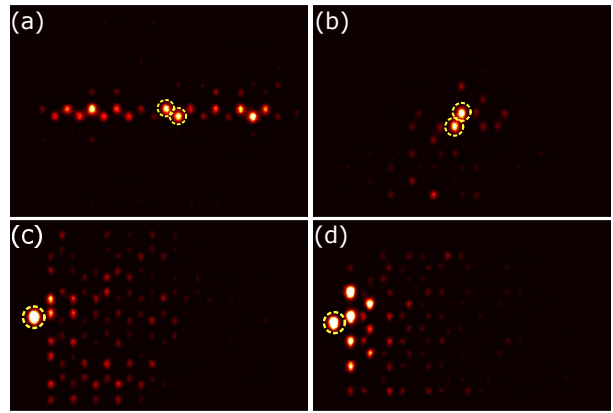}
	\caption{(a) Diffracted light measured at the output facet when we inject directly at the center of the zig-zag and (b) the armchair domain walls. (c) Images of diffracted light measured at the output facet when $\Delta n_{\textrm{A}}$=$\Delta n_{\textrm{B}}$=$\Delta n_{\textrm{s}}$ and the straw waveguide mode is initially excited for the zig-zag orientation and (d) the armchair domain walls.  The lack of detuning leads to the lack of an edge state.  All measurements are carried out at $\lambda$=1450nm. Waveguides where light is initially injected are marked with yellow dashed circle.}
	\label{Fig_04}
\end{figure}

In order to confirm that the small edge intensity ratio and penetration ratio measured at $\lambda$=1450nm is indeed the consequence of a large edge state bandgap, as opposed to simply weak inter-waveguide coupling, we injected light at the center of the domain wall such that edge states are directly excited (Fig. \ref{Fig_04}(a) and (b)) - in other words, we did not attempt to control the energy by using the straw.
If the small penetration ratios were the result of weak coupling strength between the nearest-neighbor waveguides, the injected light would be expected to be strongly confined at the center of the waveguide array, where it is initially injected.
However, for both waveguide arrays with zig-zag and armchair domain walls, we observed light diffracting along the domain wall and into the bulk.
There is significantly more diffraction along the zig-zag edge as compared to the armchair edge because the armchair edge band is nearly flat and the zig-zag edge band is highly dispersive.
However, in both cases, there is clear diffraction into the bulk of the structure, as is expected when we do not drive at a fixed energy using the straw.
Furthermore, we examine the case where the system has no inversion breaking whatsoever, namely $\Delta n_{\textrm{A}}$=$\Delta n_{\textrm{B}}$=$\Delta n_{\textrm{s}}$.
In this case, there is no bandgap and therefore no edge state of any kind.
Upon injecting light into the straw waveguide, we observe strong diffraction into the bulk for both structures shown in Fig. \ref{Fig_01}(a,b) of the zig-zag and armchair orientation (Fig. \ref{Fig_04}(c,d)).
Taken together, these results show that the straw waveguide acts as a reliable `spectroscopy tool' for directly observing the presence in the valley-Hall edge states in the wavelength range 1450nm-1650nm.    

In summary, we have realized the photonic valley-Hall topological edge states in two-dimensional honeycomb photonic lattices with broken inversion symmetry.
We have experimentally demonstrated that it is possible to open very large bandgaps and therefore enter a fully gapped regime even for the structure with zig-zag edge domain walls, which was not possible in solid-state two-dimensional materials.
Auxiliary straw waveguides placed at either end of the domain walls made it possible to access and excite a desired energy within the bulk bandgap, allowing for a convenient `spectroscopy' tool for the waveguide array energies.  Being a time-reversal invariant system, the valley-Hall effect could provide a straightforward route towards realizing photonic topological edge states, particularly in an on-chip platform.  Thus, while valley-Hall edge states are not rigorously protected against any class of disorder, they will be protected against disorder that is sufficiently smooth (and thus does not allow inter-valley scattering).  The linear, static, and non-magnetic nature of the design will also allow for lower optical loss compared to other approaches to topologically-protected photonic states (for example, magnetic materials are typically very lossy).  Furthermore, the photonic valley-Hall effect could provide a natural platform for photonic quantum simulation of topological phenomena, perhaps by coupling the photonic modes to atoms or excitons.

During the writing of the manuscript, we became aware of an analogous work in the microwave regime \cite{Wu2017}.

\begin{acknowledgments}
M.C.R. acknowledges the National Science Foundation under award number ECCS-1509546 and the Penn State MRSEC, Center for Nanoscale Science, under award number NSF DMR-1420620 as well as the Alfred P. Sloan Foundation under fellowship number FG-2016-6418. K.P.C. acknowledges the National Science Foundation under award numbers ECCS-1509199 and DMS-1620218. 
\end{acknowledgments}

\bibliography{reference}

\end{document}